\documentclass[aps,prd,twocolumn,superscriptaddress,showpacs]{revtex4}

\usepackage{graphicx}  % needed for figures
\usepackage{dcolumn}   % needed for some tables
\usepackage{bm}        % for math
\usepackage{amssymb}   % for math

\newcommand{\met}{$E_{T}\hspace{-1.1em}/$\hspace{0.7em}}
\newcommand{\metb}{$E_{T}\hspace{-1.1em}/$\hspace{1.0em}}
\newcommand{\metc}{E_{T}\hspace{-1.1em}/\hspace{0.6em}}
\newcommand{\vmet}{$\vec{E_{T}}\hspace{-1.1em}/$\hspace{0.7em}}
\newcommand{\vmetb}{$\vec{E_{T}}\hspace{-1.1em}/$\hspace{1.0em}}
\newcommand{\vmetc}{\vec{E_{T}}\hspace{-1.1em}/\hspace{0.6em}}

\newcommand{\sstop} {\tilde{t}}
\newcommand{\ssbottom} {\tilde{b}}
\newcommand{\schi} {\tilde{\chi}}

\begin{document}

%\linenumbers
%\runninglinenumbers
%%%\pagewiselinenumbers

%%%\begin{flushright}
%%%CDF/PUB/EXOTIC/PUBLIC/8781 \\
%%%\end{flushright}

\hyphenation{simu-la-tion}
\title{\begin{boldmath}
Searches for Direct Pair Production of Supersymmetric Top and Supersymmetric Bottom Quarks
in $p\bar{p}$ Collisions at $\sqrt{s}=1.96$~TeV
\end{boldmath}}
%%%%\author{\input{Default_Run2_Auth.tex}}

\affiliation{Institute of Physics, Academia Sinica, Taipei, Taiwan 11529, Republic of China} 
\affiliation{Argonne National Laboratory, Argonne, Illinois 60439} 
\affiliation{Institut de Fisica d'Altes Energies, Universitat Autonoma de Barcelona, E-08193, Bellaterra (Barcelona), Spain} 
\affiliation{Baylor University, Waco, Texas  76798} 
\affiliation{Istituto Nazionale di Fisica Nucleare, University of Bologna, I-40127 Bologna, Italy} 
\affiliation{Brandeis University, Waltham, Massachusetts 02254} 
\affiliation{University of California, Davis, Davis, California  95616} 
\affiliation{University of California, Los Angeles, Los Angeles, California  90024} 
\affiliation{University of California, San Diego, La Jolla, California  92093} 
\affiliation{University of California, Santa Barbara, Santa Barbara, California 93106} 
\affiliation{Instituto de Fisica de Cantabria, CSIC-University of Cantabria, 39005 Santander, Spain} 
\affiliation{Carnegie Mellon University, Pittsburgh, PA  15213} 
\affiliation{Enrico Fermi Institute, University of Chicago, Chicago, Illinois 60637} 
\affiliation{Comenius University, 842 48 Bratislava, Slovakia; Institute of Experimental Physics, 040 01 Kosice, Slovakia} 
\affiliation{Joint Institute for Nuclear Research, RU-141980 Dubna, Russia} 
\affiliation{Duke University, Durham, North Carolina  27708} 
\affiliation{Fermi National Accelerator Laboratory, Batavia, Illinois 60510} 
\affiliation{University of Florida, Gainesville, Florida  32611} 
\affiliation{Laboratori Nazionali di Frascati, Istituto Nazionale di Fisica Nucleare, I-00044 Frascati, Italy} 
\affiliation{University of Geneva, CH-1211 Geneva 4, Switzerland} 
\affiliation{Glasgow University, Glasgow G12 8QQ, United Kingdom} 
\affiliation{Harvard University, Cambridge, Massachusetts 02138} 
\affiliation{Division of High Energy Physics, Department of Physics, University of Helsinki and Helsinki Institute of Physics, FIN-00014, Helsinki, Finland} 
\affiliation{University of Illinois, Urbana, Illinois 61801} 
\affiliation{The Johns Hopkins University, Baltimore, Maryland 21218} 
\affiliation{Institut f\"{u}r Experimentelle Kernphysik, Universit\"{a}t Karlsruhe, 76128 Karlsruhe, Germany} 
\affiliation{High Energy Accelerator Research Organization (KEK), Tsukuba, Ibaraki 305, Japan} 
\affiliation{Center for High Energy Physics: Kyungpook National University, Taegu 702-701, Korea; Seoul National University, Seoul 151-742, Korea; SungKyunKwan University, Suwon 440-746, Korea} 
\affiliation{Ernest Orlando Lawrence Berkeley National Laboratory, Berkeley, California 94720} 
\affiliation{University of Liverpool, Liverpool L69 7ZE, United Kingdom} 
\affiliation{University College London, London WC1E 6BT, United Kingdom} 
\affiliation{Centro de Investigaciones Energeticas Medioambientales y Tecnologicas, E-28040 Madrid, Spain} 
\affiliation{Massachusetts Institute of Technology, Cambridge, Massachusetts  02139} 
\affiliation{Institute of Particle Physics: McGill University, Montr\'{e}al, Canada H3A~2T8; and University of Toronto, Toronto, Canada M5S~1A7} 
\affiliation{University of Michigan, Ann Arbor, Michigan 48109} 
\affiliation{Michigan State University, East Lansing, Michigan  48824} 
\affiliation{University of New Mexico, Albuquerque, New Mexico 87131} 
\affiliation{Northwestern University, Evanston, Illinois  60208} 
\affiliation{The Ohio State University, Columbus, Ohio  43210} 
\affiliation{Okayama University, Okayama 700-8530, Japan} 
\affiliation{Osaka City University, Osaka 588, Japan} 
\affiliation{University of Oxford, Oxford OX1 3RH, United Kingdom} 
\affiliation{University of Padova, Istituto Nazionale di Fisica Nucleare, Sezione di Padova-Trento, I-35131 Padova, Italy} 
\affiliation{LPNHE, Universite Pierre et Marie Curie/IN2P3-CNRS, UMR7585, Paris, F-75252 France} 
\affiliation{University of Pennsylvania, Philadelphia, Pennsylvania 19104} 
\affiliation{Istituto Nazionale di Fisica Nucleare Pisa, Universities of Pisa, Siena and Scuola Normale Superiore, I-56127 Pisa, Italy} 
\affiliation{University of Pittsburgh, Pittsburgh, Pennsylvania 15260} 
\affiliation{Purdue University, West Lafayette, Indiana 47907} 
\affiliation{University of Rochester, Rochester, New York 14627} 
\affiliation{The Rockefeller University, New York, New York 10021} 
\affiliation{Istituto Nazionale di Fisica Nucleare, Sezione di Roma 1, University of Rome ``La Sapienza," I-00185 Roma, Italy} 
\affiliation{Rutgers University, Piscataway, New Jersey 08855} 
\affiliation{Texas A\&M University, College Station, Texas 77843} 
\affiliation{Istituto Nazionale di Fisica Nucleare, University of Trieste/\ Udine, Italy} 
\affiliation{University of Tsukuba, Tsukuba, Ibaraki 305, Japan} 
\affiliation{Tufts University, Medford, Massachusetts 02155} 
\affiliation{Waseda University, Tokyo 169, Japan} 
\affiliation{Wayne State University, Detroit, Michigan  48201} 
\affiliation{University of Wisconsin, Madison, Wisconsin 53706} 
\affiliation{Yale University, New Haven, Connecticut 06520} 
\author{T.~Aaltonen}
\affiliation{Division of High Energy Physics, Department of Physics, University of Helsinki and Helsinki Institute of Physics, FIN-00014, Helsinki, Finland}
\author{A.~Abulencia}
\affiliation{University of Illinois, Urbana, Illinois 61801}
\author{J.~Adelman}
\affiliation{Enrico Fermi Institute, University of Chicago, Chicago, Illinois 60637}
\author{T.~Affolder}
\affiliation{University of California, Santa Barbara, Santa Barbara, California 93106}
\author{T.~Akimoto}
\affiliation{University of Tsukuba, Tsukuba, Ibaraki 305, Japan}
\author{M.G.~Albrow}
\affiliation{Fermi National Accelerator Laboratory, Batavia, Illinois 60510}
\author{S.~Amerio}
\affiliation{University of Padova, Istituto Nazionale di Fisica Nucleare, Sezione di Padova-Trento, I-35131 Padova, Italy}
\author{D.~Amidei}
\affiliation{University of Michigan, Ann Arbor, Michigan 48109}
\author{A.~Anastassov}
\affiliation{Rutgers University, Piscataway, New Jersey 08855}
\author{K.~Anikeev}
\affiliation{Fermi National Accelerator Laboratory, Batavia, Illinois 60510}
% ...
\collaboration{CDF Collaboration\footnote{With visitors from $^a$University of Athens, 15784 Athens, Greece, 
$^b$University of Bristol, Bristol BS8 1TL, United Kingdom, 
$^c$University Libre de Bruxelles, B-1050 Brussels, Belgium, 
$^d$Cornell University, Ithaca, NY  14853, 
$^e$University of Cyprus, Nicosia CY-1678, Cyprus, 
$^f$University College Dublin, Dublin 4, Ireland, 
$^g$University of Edinburgh, Edinburgh EH9 3JZ, United Kingdom, 
$^h$University of Heidelberg, D-69120 Heidelberg, Germany, 
$^i$Universidad Iberoamericana, Mexico D.F., Mexico, 
$^j$University of Manchester, Manchester M13 9PL, England, 
$^k$Nagasaki Institute of Applied Science, Nagasaki, Japan, 
$^l$University de Oviedo, E-33007 Oviedo, Spain, 
$^m$University of London, Queen Mary College, London, E1 4NS, England, 
$^n$University of California Santa Cruz, Santa Cruz, CA  95064, 
$^o$Texas Tech University, Lubbock, TX  79409, 
$^p$University of California, Irvine, Irvine, CA  92697, 
$^q$IFIC(CSIC-Universitat de Valencia), 46071 Valencia, Spain, 
}}
\noaffiliation

%\author{\input{July_2006_Authors.tex}}
%\affiliation{\input{affilation.tex}}

\date{\today}
\begin{abstract}
We search for direct pair production of
supersymmetric top quarks and supersymmetric bottom quarks
in proton-antiproton collisions at $\sqrt{s} = 1.96$ TeV,
using 295~pb$^{-1}$ of data recorded by the Collider Detector at Fermilab
(CDF II) experiment.
The supersymmetric top (supersymmetric bottom) quarks are selected by
reconstructing their decay into
a charm (bottom) quark and a neutralino, which is assumed to be the
lightest supersymmetric particle.
The signature of such processes is two energetic heavy-flavor jets
and missing transverse energy.
The number of events that pass our selection for each search
process is consistent with the expected standard model background.
By comparing our results to the theoretical production
cross sections of the supersymmetric top and
supersymmetric bottom quarks in the minimal supersymmetric standard model,
we exclude, at a 95\% confidence level in the frame of that model,
a supersymmetric top quark mass up to 132 GeV/$c^{2}$ for a neutralino mass of 48 GeV/$c^{2}$,
and a supersymmetric bottom quark mass up to 193 GeV/$c^{2}$
for a neutralino mass of 40 GeV/$c^{2}$.
\end{abstract} 
\maketitle

%================================================
\section{Introduction}
%================================================
Supersymmetry (SUSY) is an extension of the standard model (SM)
of particle physics that overcomes some of the theoretical problems in the
SM by introducing a new degree of freedom \cite{SUSY}.
In this model a bosonic supersymmetric partner is assigned to every SM fermion helicity state,
and a fermionic superpartner to every SM boson.
Thus, the SM quark helicity states $q_{L}$ and $q_{R}$ acquire scalar partners
$\tilde{q}_{L}$ and $\tilde{q}_{R}$.
The mass eigenstates of each supersymmetric quark (squark) can be
a mixture of their weak eigenstates, quantified by a mixing angle $\theta$.
The difference in the mass eigenvalues depends on several factors.
In the case of the supersymmetric top quark (stop), due to the large top quark mass and the
large value of its Yukawa coupling constant (Higgs-to-top coupling),
there can be a significant difference in the mass between the two mass eigenstates
$\sstop_{1}$ and $\sstop_{2}$.
In the case of the supersymmetric bottom quark (sbottom),
a large mass difference between the two mass eigenstates
$\ssbottom_{1}$ and $\ssbottom_{2}$ can occur if the ratio of the vacuum expectation values of
the two Higgs fields expected in SUSY is large \cite{SMartin}.
In both cases it is likely that the less massive
stop ($\sstop_{1}$) and sbottom ($\ssbottom_{1}$)
could be lighter than the first two generations of supersymmetric quarks.
In fact the $\sstop_{1}$ could be lighter than the top quark if the top Yukawa coupling strength
or the stop mixing is strong enough. The requirement of a light stop is also
a feature of many baryogenesis models \cite{baryogenesis}.

In this paper, we describe two analyses searching for stop and sbottom
production at the Tevatron using a data sample of 295 pb$^{-1}$ integrated
luminosity collected with the CDF II detector. Both analyses are performed within
the $R$-Parity \cite{Rparity} conserving minimal supersymmetric standard model framework.
A consequence of $R$-Parity conservation is that all SUSY particles are pair produced,
and the lightest SUSY particle (LSP) is stable.
If the LSP interacts weakly it is a good candidate for cold dark matter
and it escapes our detection.
%%%Cosmological contraints require the LSP to be neutral and colorless \cite{SUSYLSP}.
%%%Thus, the LSP would interact only weakly and escapes detection, which makes
%%%it a good candidate for cold dark matter.

\begin{figure}[htbp]
%%%\centering
\includegraphics[width=9cm]{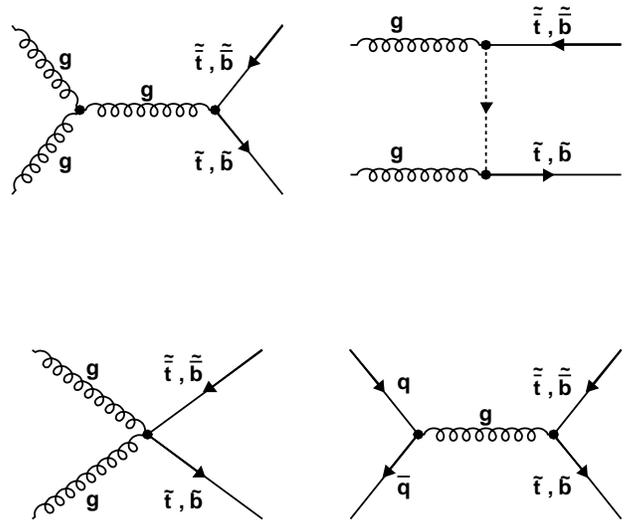}
\caption{Leading-order Feynman diagrams for pair production of
         stop and sbottom at the Tevatron.}
\label{Fig:stsbprod_1}
\end{figure}

At the Tevatron, stop and sbottom are expected to be produced in pairs
mainly via $gg$ fusion and $q\bar{q}$ annihilation, as shown in Fig. \ref{Fig:stsbprod_1}.
At leading order, their production cross sections depend essentially only on their masses.
For a center-of-mass energy of 1.96~TeV, the next-to-leading order (NLO) cross section,
calculated with {\sc prospino} \cite{prospino}, ranges from 50.3 pb to 0.25 pb
for stop and sbottom masses from 80 GeV/$c^{2}$ to 200 GeV/$c^{2}$.
In the calculation, the renormalization and factorization scales are set to the mass of
stop or sbottom ($Q_{\mbox{rf}}=m_{\sstop_{1},\ssbottom_{1}}$),
and the parton distribution functions (PDFs) are from {\sc cteq6m} \cite{CTEQ6M}.

%%%%%%%%%%%%%%%%%%%%%%%%%%%%%%%%%%%%
%%                                 %
%%%%%%%%%%%%%%%%%%%%%%%%%%%%%%%%%%%%

\begin{figure}[htbp]
\centering
\includegraphics[width=9.5cm]{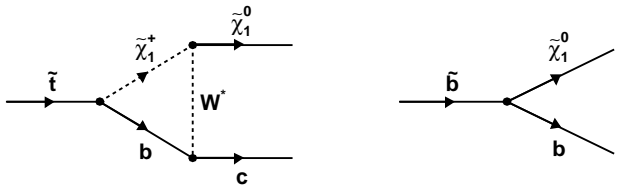}
\caption{The decay channels of stop and sbottom considered in this paper.}
\label{Fig:stsbdecay_1}
\end{figure}

The stop and sbottom can decay in many channels, depending on the mass difference
between stop/sbottom and other SUSY and SM particles.
Here we consider the SUSY parameter space where the stop and sbottom
are relatively light.
In the case of stop, the flavor changing loop decay $\sstop_{1} \rightarrow c\schi^{0}_{1}$
dominates if $m_{\sstop_{1}} > m_{c} + m_{\schi^{0}_{1}}$, but
$m_{\sstop_{1}} < m_{b} + m_{\schi^{\pm}_{1}}$
and $m_{\sstop_{1}} < m_{W} + m_{b} + m_{\schi^{0}_{1}}$.
In the sbottom search, the $\ssbottom_{1} \rightarrow b\schi^{0}_{1}$
is the only relevant decay if
 $m_{\ssbottom_{1}} > m_{b} + m_{\schi^{0}_{1}}$, but
$m_{\ssbottom_{1}} < m_{b} + m_{\schi^{0}_{2}}$ and
$m_{\ssbottom_{1}} < m_{t} + m_{\schi^{\pm}_{1}}$.
The neutralinos ($\schi^{0}_{1,2,3,4}$) and the charginos ($\schi^{\pm}_{1,2}$)
are the SUSY partners of the electroweak bosons
and are labelled in order of increasing mass.
Therefore, we search for the processes
$p\bar{p} \rightarrow \sstop_{1} \bar{\sstop_{1}}
 \rightarrow (c\schi^{0}_{1}) (\bar{c}\schi^{0}_{1})$,
and
$p\bar{p} \rightarrow \ssbottom_{1} \bar{\ssbottom_{1}} \rightarrow (b\schi^{0}_{1})
 (\bar{b}\schi^{0}_{1})$,
as shown in Fig. \ref{Fig:stsbdecay_1}.
We assume $\schi^{0}_{1}$ is the LSP.
Thus, the experimental signature for stop and sbottom pair production processes
is a pair of acollinear heavy-flavor jets (ie. $b$ and $c$ jets),
and large missing transverse energy (\met) coming from the escaping LSPs.
The searches assume that both stop and sbottom decay very close to the interaction point.
For $m_{\sstop_{1}}=100$ GeV/$c^{2}$ ($m_{\ssbottom_{1}}=160$ GeV/$c^{2}$) and
$m_{\schi^{0}_{1}}=60$ GeV/$c^{2}$, the expected lifetime of stop (sbottom) with only the decay
$\sstop_{1} \rightarrow c\schi^{0}_{1}$
($\ssbottom_{1} \rightarrow b\schi^{0}_{1}$) is of the order of $10^{-15}$
($10^{-23}$) seconds which corresponds to a natural decay length of 0.3
$\mu$m ($3\times 10^{-9}$ $\mu$m) \cite{Stop_Width,Sbottom_Width}.

%%%%%%%%%%%%%%%%%%%%%%%%%%%%%%%%%%%%
%%                                 %
%%%%%%%%%%%%%%%%%%%%%%%%%%%%%%%%%%%%

Previous searches for stop and sbottom have been performed at
LEP and at the Tevatron
\cite{LEPStopSbottom,CDFRun1StopSbottom,D0Run1Stop,blsneu,stopRPV,Topstop,
bbbb,D0Run2Stop,D0Run2Sbottom}.
For the search topology studied in the present
analysis, LEP excludes  stop (sbottom) masses smaller than
$\approx 100$ GeV/$c^{2}$ ($\approx 100$ GeV/$c^{2}$), independent of the
difference between stop (sbottom) and neutralino $\schi^{0}_{1}$ masses
\cite{LEPStopSbottom}. Recent results from the D\O\ Collaboration in the same topology
\cite{D0Run2Stop,D0Run2Sbottom}, based on Run II data, have extended Tevatron's
Run I reach \cite{CDFRun1StopSbottom,D0Run1Stop}
by excluding stop masses up to $\sim 133$ GeV/$c^{2}$,
and sbottom masses up to $\sim 220$ GeV/$c^{2}$.
These results are also shown in Fig. \ref{Fig:mass_plane_exclude_stop} and
\ref{Fig:mass_plane_exclude_sbottom} of this paper.

%================================================
\section{The Detector and Data Sample}
%================================================
CDF II is  a general-purpose detector that is described in detail
elsewhere \cite{cdf}. The components relevant to this analysis are briefly described here.
The charged-particle tracking system is closest to the beam pipe,
and consists of multiple layers of silicon micro-strip detectors,
which cover a pseudorapidity region $|\eta|< 2$,
and a large open-cell drift chamber covering the pseudorapidity region
$|\eta|< 1$ \cite{svx,cot,eta}.
The silicon microstrips of the silicon detectors have a pitch of 25 to 65 $\mu$m,
depending on the layer, thus allowing a precise measurement
of a track's impact parameter with respect to the primary vertex.
The tracking  system  is enclosed in a superconducting solenoid,
which in turn is surrounded by calorimeters.
The calorimeter system \cite{cal} is organized into electromagnetic
and hadronic sections segmented in projective tower geometry, and covers  the region $|\eta|< 3.6$. 
The electromagnetic calorimeters use lead-scintillator sampling,
whereas the hadron calorimeters use iron-scintillator sampling construction.
The transverse energy resolution of the electromagnetic calorimeters is
$\sigma(E_{T})/E_{T}=\frac{13.5\%}{\sqrt{E_{T}{\rm(GeV)}}} \oplus 2\%$ for the central region ($|\eta|<1$),
and $\sigma(E)/E=\frac{16\%}{\sqrt{E}{\rm(GeV)}} \oplus 1\%$ for the forward region ($|\eta|>1$).
The transverse energy resolution of the hadronic calorimeters is
$\sigma(E_{T})/E_{T}=\frac{75\%}{\sqrt{E_{T}{\rm(GeV)}}} \oplus 3\%$ for the central region,
and $\sigma(E)/E=\frac{80\%}{\sqrt{E}{\rm(GeV)}} \oplus 5\%$ for the forward region.
The present analysis exploits the information of the central muon system,
which is located outside of the calorimeter and covers the range $|\eta|<1$.

The data sample for this analysis was collected using a \met+jets trigger,
which is implemented in three levels of online event selection. 
The \metb is defined as the energy imbalance in the plane transverse
to the beam direction \cite{eta},
and a jet is defined as a localized energy deposition in the calorimeter.
In the first and second levels of the trigger, \metb is required  to be
greater than $25$~GeV and is calculated by summing over calorimeter
trigger towers with transverse energies above 1~GeV.
In the second level there must be at least two jets
with $E_{T}>10$ GeV.
In the third level, \metb is recalculated using the full calorimeter segmentation with
a tower energy threshold of  100~MeV and is required  to be greater than $35$~GeV.

In the offline processing, jets are reconstructed from the calorimeter towers
using a cone algorithm with fixed radius $\Delta R \equiv\sqrt{\Delta\eta^2+\Delta\phi^2}=0.4$
in $\eta-\phi$ space \cite{jetalg}.
The jet $E_{T}$ measurements and \metb are corrected for detector effects \cite{jetcorr}.

A fraction of events passing the trigger is not from $p\bar{p}$ collisions,
but from beam halo and cosmic ray sources.
To remove these events we examine the event electromagnetic fraction $F_{em}$
and charged fraction $F_{ch}$.
The $F_{em}$ is the ratio of the energy measured by the electromagnetic calorimeter to the
total energy contained in jets of cone radius $\Delta R=0.4$  with $E_{T}>10$~GeV and  $|\eta|<3.6$. 
The $F_{ch}$ is the fraction of the jet energy carried by measured charged-particle tracks ($p_T>0.5$ GeV$/c$)
averaged over jets with  $|\eta|<0.9$.
The beam halo travels parallel to the beam axis and deposits
most of its energy in the electromagnetic section of the calorimeter or in the hadronic section.
By requiring $F_{em}>0.1$ we reject events which contain little energy
in the electromagnetic section of the calorimeter.
A cosmic ray traversing the CDF detector can deposit energy in the calorimeter
without registering a track in the tracking detectors.
If the beam halo background events and cosmic ray background events, as described above,
do not overlap with beam crossing events that produce hard $p\bar{p}$ collisions,
then there will be little activity in the tracking detectors.
Therefore, by requiring $F_{ch}>0.1$ we reject backgrounds that have
little tracking activity.
More detailed explanations of these two variables and how they reduce the
non $p\bar{p}$ collision events are described in \cite{FemFch_Run1}.
We also reject events if the reconstruction cone of any jet in the event
enters an uninstrumented region of the calorimeter.

%================================================
\section{Background Sources and Event Selection}
%================================================
The dominant backgrounds to the stop and sbottom searches in the jets and
\metb signature are production of multi-jet, $W$ or $Z$ boson with jets,
single top, $t\bar{t}$, and di-boson ($WW$/$WZ$/$ZZ$) final states. 
We use the {\sc alpgen} generator to simulate the $W$ and $Z$
boson plus parton production, with  {\sc herwig} used to model parton 
showers \cite{alpgen,herwig}.
Multi-jet, top quark, and di-boson production are simulated with {\sc pythia} \cite{pythia}.
All the Monte Carlo (MC) SM background samples were generated using the {\sc cteq5l} PDFs \cite{cteq5l}.
In generating the multi-jet sample, we select events where there are
$b$ or $c$ outgoing partons/hadrons in the final state [heavy-flavor (HF) multi-jet].
To normalize the SM top background samples, we use the NLO
cross section values for single top and $t\bar{t}$ production \cite{ttbarxs,toptch,topsch}.
We use the {\sc mcfm} program to obtain the NLO cross sections for $W/Z+$jets
and di-boson production \cite{mcfm1,mcfm2}.
The HF multi-jet sample is normalized using kinematic regions in the data that are
dominated by multi-jet production.
In multi-jet production events, the \metb is usually due to jet energy mis-measurement.
In this case the \vmetb tends to point in the same direction as the jet whose
energy is mis-measured.
In the \met+jets data sample used in this analysis, the events at higher \metb
are dominated by non-multi-jet SM contributions.
Therefore, the multi-jet dominant kinematic regions are low \metb regions ($50 < \metc < 70$ GeV),
and regions where a jet is aligned in the direction of \met.
We obtain an average normalization factor
$k_{HF multi-jet}=1.46\pm0.37$ where the uncertainty is mainly due to
the uncertainty from jet energy calibration and resolution \cite{jetcorr}.
The $k_{HF multi-jet}$ factor is used to normalize the HF multi-jet sample
to estimate the HF multi-jet contribution in the signal region.

Data selection is optimized by maximizing the statistical significance of a simulated
stop/sbottom signal over the expected background events in the data.
The optimization is performed prior to examining the signal regions of the data.
As the signal production cross section and event kinematics (for example jet $E_{T}$, \metb)
could vary significantly across stop and sbottom masses,
we determine separate sets of optimized cuts for
three mass ranges. 
The ``low'', ``medium'', and ``high'' mass ranges used for the stop (sbottom) search are
$m_{\sstop_{1}}<100$ GeV/$c^{2}$, $100 \leq m_{\sstop_{1}} < 120$ GeV/$c^{2}$,
and $m_{\sstop_{1}} \geq 120$ GeV/$c^{2}$
($m_{\ssbottom_{1}}<140$ GeV/$c^{2}$, $140 \leq m_{\ssbottom_{1}} < 180$ GeV/$c^{2}$,
and $m_{\ssbottom_{1}} \geq 180$ GeV/$c^{2}$).
The selection cuts for the stop and sbottom are summarized in
Table \ref{tab:Evt_Select_Cuts_stop} and  \ref{tab:Evt_Select_Cuts_sbottom} respectively.
At the initial stage of event selection, the data sample is dominated by
multi-jet background events. We employ several selection cuts to
minimize their contribution.
To reduce multi-jet background and avoid regions where the \metb trigger is
inefficient, we require $\metc>50$~GeV.
The \metb trigger is $\sim 75\%$ efficient in this region.
Next we require that there be only two or three reconstructed jets in $|\eta|<2$.
Events with any additional jets with  $E_{T}> 8$~GeV and $2<|\eta|<3.6$
are rejected.
Most of the time the two highest-$E_{T}$ jets in multi-jet events
are anti-parallel. However, this is not the case in the stop (sbottom)
pair production since the $c$ ($b$) jet recoils against the $\schi^{0}_{1}$ in 
the stop (sbottom) decay.
Therefore, we require that the opening angle between the two highest $E_{T}$ jets
be less than 160 degrees in the plane perpendicular to the beam.
The large \metb in multi-jet events that survive the earlier cuts
is usually due to jet energy mis-measurement.
Thus, to further reduce this background we require a minimum azimuthal separation between
the direction of the jets and \vmetb of $\Delta\phi({\rm{Jet}},\metc)>45^{\circ}$.
For multi-jet events that pass the minimum azimuthal separation requirement,
but have large \metb due to jet energy mis-measurement, the magnitudes of the
\metb and the second leading jet's transverse energy are often anti-correlated.
Therefore, we require the sum $E_{T}(\rm{Jet2})+$\metb to be above
the values listed in Table \ref{tab:Evt_Select_Cuts_stop} and \ref{tab:Evt_Select_Cuts_sbottom}.

%%%%%%%%%%%%%%%%%%%%%%%%%%%%%%%%%%%%
% TABLE : Selection cuts (stop)    %
%%%%%%%%%%%%%%%%%%%%%%%%%%%%%%%%%%%%
\begin{table*}[b,h,t]
  \caption{The event selection cuts for pair production of stop
           in the ``low'', ``medium'', and ``high'' stop mass regions.
           Jet1, Jet2, and Jet3 are, respectively, the first, second,
           and third leading jets.
           The cuts that are listed only under the ``medium'' column
           are common for all three mass ranges.}
  \label{tab:Evt_Select_Cuts_stop}
  \begin{ruledtabular}
    \begin{tabular}{lccc}
Mass range & Low & ~Medium & High \\
(GeV/$c^{2}$)                       & $<100$ & $100-120$ & $>120$ \\ \hline
\metb (GeV)      & $>50$ & $>50$ & $>50$ \\
$E_{T}$ Jet1 (GeV) & $>35$ & $>45$ & $>55$ \\
$E_{T}$ Jet2 (GeV) & $>15$ & $>15$ & $>25$ \\
$E_{T}$ Jet5 (GeV) & $>15$ & $>15$ & $>15$ \\
$\mid \eta \mid$ : Jet1, Jet2, Jet3 & \multicolumn{3}{c}{$\mid \eta_{1} \mid < 1.2$,
							 $\mid \eta_{2} \mid < 1.5$,
							 $\mid \eta_{3} \mid < 2.0$} \\
Veto additional jets & \multicolumn{3}{c}{$E_{T}>8$ GeV, $2 < \mid \eta \mid < 3.6$} \\
$\Delta\phi(\rm{Jet1,Jet2})$ accepted range (deg) & $70 - 160$ & $70 - 160$ & $70 - 160$ \\
$\Delta\phi({\rm{Jet}},\metc)$ accepted range (deg) & $45 - 180$ & $45 - 180$ & $45 - 180$ \\
$E_{T}({\rm{Jet2}}) + \metc$ (GeV) & $>65$ & $>85$ & $>105$ \\
$E^{v}_{T_{J12MET}}$ (GeV) & $<15$ & $<15$ & $<15$ \\
Lepton Veto    & YES & YES & YES \\
minimum \# tracks in jet ($\mid \eta \mid < 1$) & 4 & 4 & 4 \\
%
%%%{\sc jet-probability} Tagging  & $>=1$ tag (\@ {\it JP} $<5\%$) & $>=1$ tag (\@ {\it JP} $<5\%$) & $>=1$ tag (\@ {\it JP} $<5\%$) \\
{\sc jet-probability} Tagging  & \multicolumn{3}{c}{$>=1$ tag (\@ {\it JP} $<5\%$) } \\
    \end{tabular}
  \end{ruledtabular}
\end{table*}

%%%%%%%%%%%%%%%%%%%%%%%%%%%%%%%%%%%%
% TABLE : Selection cuts (sbottom) %
%%%%%%%%%%%%%%%%%%%%%%%%%%%%%%%%%%%%
\begin{table*}[b,h,t]
  \caption{The event selection cuts for pair production of sbottom
           in the ``low'', ``medium'', and ``high'' sbottom mass regions.
	   Jet1, Jet2, and Jet3 are, respectively, the first, second,
           and third leading jets.
           The cuts that are listed only under the ``medium'' column
           are common for all three mass ranges.}
  \label{tab:Evt_Select_Cuts_sbottom}
  \begin{ruledtabular}
    \begin{tabular}{lccc}
Mass range & Low & Medium & High \\
(GeV/$c^{2}$)                       & $<140$ & $140-180$ & $>180$ \\ \hline
\metb (GeV)      & $>50$ & $>55$ & $>65$ \\
$E_{T}$ Jet1 (GeV) & $>35$ & $>55$ & $>75$ \\
$E_{T}$ Jet2 (GeV) & $>15$ & $>15$ & $>35$ \\
$E_{T}$ Jet3 (GeV) & $>15$ & $>15$ & $>15$ \\
$\mid \eta \mid$ : Jet1, Jet2, Jet3 &
   \multicolumn{3}{c}{$\mid \eta_{1} \mid < 1.2$,
 	              $\mid \eta_{2} \mid < 1.5$,
                      $\mid \eta_{3} \mid < 2.0$}     \\
Veto additional jets & \multicolumn{3}{c}{$E_{T}>8$ GeV,
                                            $2 < \mid \eta \mid < 3.6$} \\
$\Delta\phi(\rm{Jet1,Jet2})$ accepted range (deg) & $60 - 160$ & $50 - 160$ & $40 - 160$ \\
$\Delta\phi({\rm{Jet}},\metc)$ accepted range (deg) & $45 - 180$ & $45 - 180$ & $45 - 180$ \\
$E_{T}({\rm{Jet2}}) + \metc$ (GeV) & $>80$ & $>120$ & $>160$ \\
$E^{v}_{T_{J12MET}}$ (GeV) & $<15$ & $<15$ & $<15$ \\
Lepton Veto    & YES & YES & YES     \\
minimum \# tracks in jet ($\mid \eta \mid < 1$) & 4 & 4 & 4 \\
%
%%%{\sc jet-probability} Tagging  & $>=1$ tag (\@ {\it JP} $<1\%$) & $>=1$ tag (\@ {\it JP} $<1\%$) & $>=1$ tag (\@ {\it JP} $<1\%$) \\
{\sc jet-probability} Tagging  & \multicolumn{3}{c}{$>=1$ tag (\@ {\it JP} $<1\%$)} \\
    \end{tabular}
  \end{ruledtabular}
\end{table*}

In stop (sbottom) pair production, where each stop (sbottom)
decays into a charm (bottom) quark and a neutralino, the vector sum of the
transverse energy of the two $c$ jets ($b$ jets) should balance against the
missing transverse energy from the escaping neutralinos.
We explore the correlation between the missing transverse energy
and the transverse energy of the first and second leading jets through the variable
$E^{v}_{T_{J12MET}}$, which is defined as
\begin{equation}
  E^{v}_{T_{J12MET}} = (\vec{E}_{T_{J1}} + \vec{E}_{T_{J2}} + \vmetc)
	\cdot \Bigg(\frac{\vec{E}_{T_{J1}} + \vec{E}_{T_{J2}}}
                         {\mid \vec{E}_{T_{J1}} + \vec{E}_{T_{J2}} \mid}\Bigg).
\end{equation}
$\vec{E}_{T_{J1}}$  and $\vec{E}_{T_{J2}}$ are, respectively, the vectors
of the transverse energy of the first and second leading jets.
For pair production of stop (sbottom), $E^{v}_{T_{J12MET}}$ is expected
to be insignificant, i.e. within the calorimeter jet energy resolution.
For multi-jet production, due to its high jet multiplicity,
the missing transverse energy caused by the mis-measurement of a
single jet's energy, may not balance against the transverse energy
of the first and second leading jets.
In top-quark pair production, the missing transverse energy caused by
the escaping neutrino in the $W$ boson decay also does not necessarily
balance against the first and second leading jets in the event.
Therefore, $E^{v}_{T_{J12MET}}$ is large in these background events.
To reduce the contribution from top quark production, and to further reduce
multi-jet background, we require $E^{v}_{T_{J12MET}} < 15$ GeV 
for both stop and sbottom analyses.

To reduce the background contribution from $W/Z+$jets and top quark production, 
we reject events with one or more identified leptons.
Candidate electrons must have a track associated with a cluster in the
electromagnetic calorimeter with $E_{T} >10$~GeV.
Its electromagnetic to hadronic energy ratio and shower profile,
must also be consistent with that expected for electrons.
Candidate muons are identified as tracks with $p_{T}>10$~GeV$/c$
that extrapolate to hits in the muon chambers and to energy deposited
in the calorimeters consistent with a minimum ionizing particle.
To increase lepton detection efficiency, candidate leptons (electrons, muons, taus)
are also identified by isolated tracks with $p_{T}>10$~GeV$/c$.
The number of tracks associated with the first and second leading jets should be
four or more.
This selection reduces contributions from $W+$jets and $Z+$jets in which
the gauge bosons decay into tau leptons.

For the stop and sbottom search, the signal contains a large fraction of
heavy-flavor jets compared to the SM background, which is dominated by light-flavor
jets.
To enhance the signal over SM background, we identify the heavy-flavor jets
using the {\sc jet-probability} ({\it JP}) algorithm \cite{jetprobR2}.
Jets from heavy-flavor partons are characterized by secondary decays
that are displaced from the primary vertex; thus
their tracks have a large impact parameter.
Light-flavor jets appear to come from the primary interaction and their
tracks' impact parameter are consistent with
the primary vertex (within the resolution of the tracking detector).
The {\it JP} algorithm examines the impact parameter of each track from a candidate jet
and computes a probability that the jet is a light/heavy jet. Jets from the
primary (secondary) vertex are assigned a large (small) {\it JP} value.
For the stop (sbottom) search at least one jet was required to have
a {\it JP} $<5\%$ ({\it JP} $<1\%$).
A looser cut is used to tag $c$ jets as their lifetime is shorter
compared to $b$ jets.
The efficiency to tag a fiducial $c$ or $b$ jet increases with the
transverse energy of the jet, and plateaus at $E_{T}\sim 80$ GeV.
The average efficiency to tag a fiducial $c$ jet ($b$ jet) with {\it JP} value of
{\it JP} $<5\%$ ({\it JP} $<1\%$) is $\sim 17\%$ ($\sim 40\%$) \cite{jetprobR2}.

%================================================
\subsection{Light-Flavor Background}
%================================================
A significant source of background is that due to mis-identification of
light-flavor jets as heavy flavor.
In this case one or more light-flavor jets are tagged as a $c$ jet or $b$ jet
by the {\it JP} algorithm (mis-tag).
A detailed description of the mis-tag rate is given in \cite{jetprobR2}.
Here we provide a brief summary of its measurement and how it is applied
in the stop and sbottom searches.
The mis-tag rate is measured in inclusive jet samples.
The rate of mis-identifying a light-flavor jet as a heavy-flavor jet
is about $\sim 1\%$ ($\sim 5\%$) for tagging at {\it JP} $<1\%$ ({\it JP} $<5\%$).
Uncertainties on the mis-tag rates are due to uncertainties in the
contribution of long-lived $\Lambda$'s and K's in the light-flavor jets
and uncertainties in the effect of particle interactions in the detector
materials; these relative uncertainties are 9\% for {\it JP} $<1\%$,
and 13\% for {\it JP} $<5\%$.
An additional relative uncertainty on the mis-tag rate of $6.7\%$ ($4.7\%$)
for {\it JP} cut of {\it JP} $<1\%$ ({\it JP} $<5\%$) is estimated by comparing the observed and
predicted tag rates in different data samples [inclusive jet samples taken with
different jet $E_{T}$ thresholds, and high jet multiplicity ($\ge 4$ jets) sample].
To estimate the contribution from light-flavor background
in the final data sample, we apply the measured mis-tag rate to the data sample,
i.e. we multiply each jet by the mis-tag probability,
after all selection cuts are applied except the heavy-flavor jet tagging.
In the estimate of the heavy-flavor background contributions with MC samples,
we check for the existence of a $b$ or a $c$ parton or hadron within a cone
of $\Delta R=0.4$ around the jet before we tag the it,
to avoid double counting of background events due to light-flavor jets.
Since a looser tagging requirement is used to tag the $c$ jets in the stop search
than to tag the $b$ jets in the sbottom search, we expect the stop analysis
to have a larger fraction of mis-tag background compared to the sbottom analysis.
This can be seen in Tables \ref{tab:bgstop} and \ref{tab:bgsbottom}.

%%%%%%%%%%%%%%%%%%%%%%%%%%%%%%%%%%%%
% TABLE : SM BG (STOP)             %    
%%%%%%%%%%%%%%%%%%%%%%%%%%%%%%%%%%%%
\begin{table*}[b,h,t]
  \caption{The number of observed data events, and the number of expected events
           from standard model sources in the stop signal region.
           The first uncertainty is from limited simulation statistics and the
           second is from systematic uncertainties.
	  }
  \label{tab:bgstop}
  \begin{ruledtabular}
    \begin{tabular}{lccc}
Mass range                          & Low & ~Medium & High \\
(GeV/$c^{2}$)                       & $<100$ & ~$100-120$ & $>120$ \\ \hline
Process          & \multicolumn{3}{c}{Events expected} \\ \hline
$W+\rm{jets}$    & $11.5 \pm 2.4 \pm 2.6 $ & ~$9.3 \pm 2.3 \pm 2.1$ & ~$4.0 \pm 1.5 \pm 0.9$ \\
$Z+\rm{jets}$    & $9.9 \pm 0.5 \pm 2.0 $ & ~$7.3 \pm 0.4 \pm 1.5$ & ~$4.1 \pm 0.3 \pm 0.8$ \\
Di-boson         & $2.5 \pm 0.1 \pm 0.5 $ & ~$2.0 \pm 0.1 \pm 0.4$ & ~$0.9 \pm 0.1 \pm 0.2$ \\
Top              & $5.2 \pm 0.2 \pm 0.8 $ & ~$4.9 \pm 0.2 \pm 0.8$ & ~$3.9 \pm 0.2 \pm 0.6$ \\
HF Multi-jet     & $32.5 \pm 5.2 \pm 8.1 $ & ~$17.9 \pm 4.0 \pm 4.5$ & ~$2.6 \pm 1.5 \pm 0.6$ \\
Mistag           & $75.4 \pm 2.2 \pm 10.7 $ & ~$53.5 \pm 2.0 \pm 7.6$ & ~$27.2 \pm 1.5 \pm 3.9$ \\
Total Expected   & $137 \pm 6.2 \pm 14.6 $  & ~$94.9 \pm 5.0 \pm 9.9$ & ~$42.7 \pm 2.6 \pm 4.6$ \\
Data             & 151                       & ~108                      & ~43 \\
    \end{tabular}
  \end{ruledtabular}
\end{table*}

%%%%%%%%%%%%%%%%%%%%%%%%%%%%%%%%%%%%
% TABLE : SM BG (SBOTTOM)          %    
%%%%%%%%%%%%%%%%%%%%%%%%%%%%%%%%%%%%
\begin{table*}[b,h,t]
  \caption{The number of observed data events, and the number of expected events
           from standard model sources in the sbottom signal region.
           The first uncertainty is from limited simulation statistics and the
           second is from systematic uncertainties. The contribution from di-boson background
           is found to be negligible in the sbottom search.
	  }
  \label{tab:bgsbottom}
  \begin{ruledtabular}
    \begin{tabular}{lccc}
Mass range & Low & ~Medium & High \\
(GeV/$c^{2}$)                       & $<140$ & ~$140-180$ & $>180$ \\ \hline
Process          & \multicolumn{3}{c}{Events expected} \\ \hline
$W+\rm{jets}$    & $5.5 \pm 1.2 \pm 1.2 $ & ~$1.5 \pm 0.6 \pm 0.3$ & ~$0.5 \pm 0.4 \pm 0.1$ \\
$Z+\rm{jets}$    & $6.0 \pm 0.4 \pm 1.1 $ & ~$2.7 \pm 0.3 \pm 0.5$ & ~$1.0 \pm 0.2 \pm 0.2$ \\
Top              & $4.2 \pm 0.2 \pm 0.6 $ & ~$2.9 \pm 0.1 \pm 0.4$ & ~$1.2 \pm 0.1 \pm 0.2$ \\
HF Multi-jet     & $18.8 \pm 4.0 \pm 4.7 $ & ~$2.6 \pm 1.5 \pm 0.7$ & ~$0 ^{+2.1}_{-0}$ \\
Mistag           & $20.5 \pm 0.6 \pm 2.3 $ & ~$8.1 \pm 0.5 \pm 0.9$ & ~$2.0 \pm 0.3 \pm 0.2$ \\
Total Expected   & $55.0 \pm 4.2 \pm 5.9 $ & ~$17.8 \pm 1.7 \pm 1.6$ & ~$4.7 ^{+2.1}_{-0.5} \pm 0.5$\\
Data             & 60                         & ~18                       & 3 \\
    \end{tabular}
  \end{ruledtabular}
\end{table*}

%================================================
\section{Detection Efficiency and Systematic Uncertainties}
%================================================
The total detection efficiencies for the stop and sbottom
signals are estimated using the {\sc pythia} event generator
and the CDF detector simulation program.
The samples were generated using the {\sc cteq5l} 
PDFs, with the renormalization
and factorization scale set to the mass of the squark in the search \cite{cteq5l}.
The total stop (sbottom) signal selection efficiency in the accessible mass region 
varies from 0.1\% to 3.4\% (0.17\% to 8.5\%).
The efficiency increases for higher stop (sbottom) mass and larger mass difference between
$\sstop_{1}$ ($\ssbottom_{1}$) and $\schi^{0}_{1}$.

We have estimated the main contributions to the systematic uncertainty on
the signal acceptance and the SM background estimation.
The uncertainty due to the jet energy scale for the SM background estimation
is of the order of 10\%,
whereas it varies from 4\% to 20\% for the signal efficiency,
depending on the stop and sbottom masses (larger uncertainty for
smaller squark mass).
The systematic uncertainty from NLO cross sections in the SM background
varies between 8\% to 13\%.
The systematic uncertainty on the efficiency for tagging
a $c$ jet ($b$ jet) is 12\% (8.6\%) \cite{jetprobR2}.
The uncertainty on the signal acceptance due to modeling of gluon radiation
from the initial-state or final-state partons is 5\%.
This is evaluated from signal MC samples generated with different levels
of initial and final state radiation.
The uncertainty on the signal efficiency due to the PDF choice is determined
to be 2\%, using the {\sc cteq6m} uncertainty PDF set.
The uncertainty from MC statistics reaches in the most selective search region
50\% for the SM background, and 10\% for signal.
The uncertainty on the trigger efficiency is 5\%,
and the uncertainty on the luminosity of the data sample is 6\% \cite{lum}.
The quoted uncertainties are relative to the estimated signal and backgrounds.
%
%

%================================================
\section{Results and Conclusions}
%================================================
The SM contributions and the total number of observed data events are shown
in Tables \ref{tab:bgstop} and \ref{tab:bgsbottom} for the stop and
sbottom searches respectively.
We find that after applying all the selection cuts,
the number of observed events are consistent with
the number of expected SM background events
for both stop and sbottom searches.
In these two searches, and for all the mass ranges,
the largest source of background events is due to mis-identification
of light-flavor jets
as $c$ or $b$ jets. The HF multi-jet background is the second largest
source of SM background in the ``low'' mass range. However, its contribution is
largely suppressed by the tighter cuts employed in the ``medium'' and ``high'' mass ranges.

\begin{figure}[htbp]
\centering
\includegraphics[width=9cm]{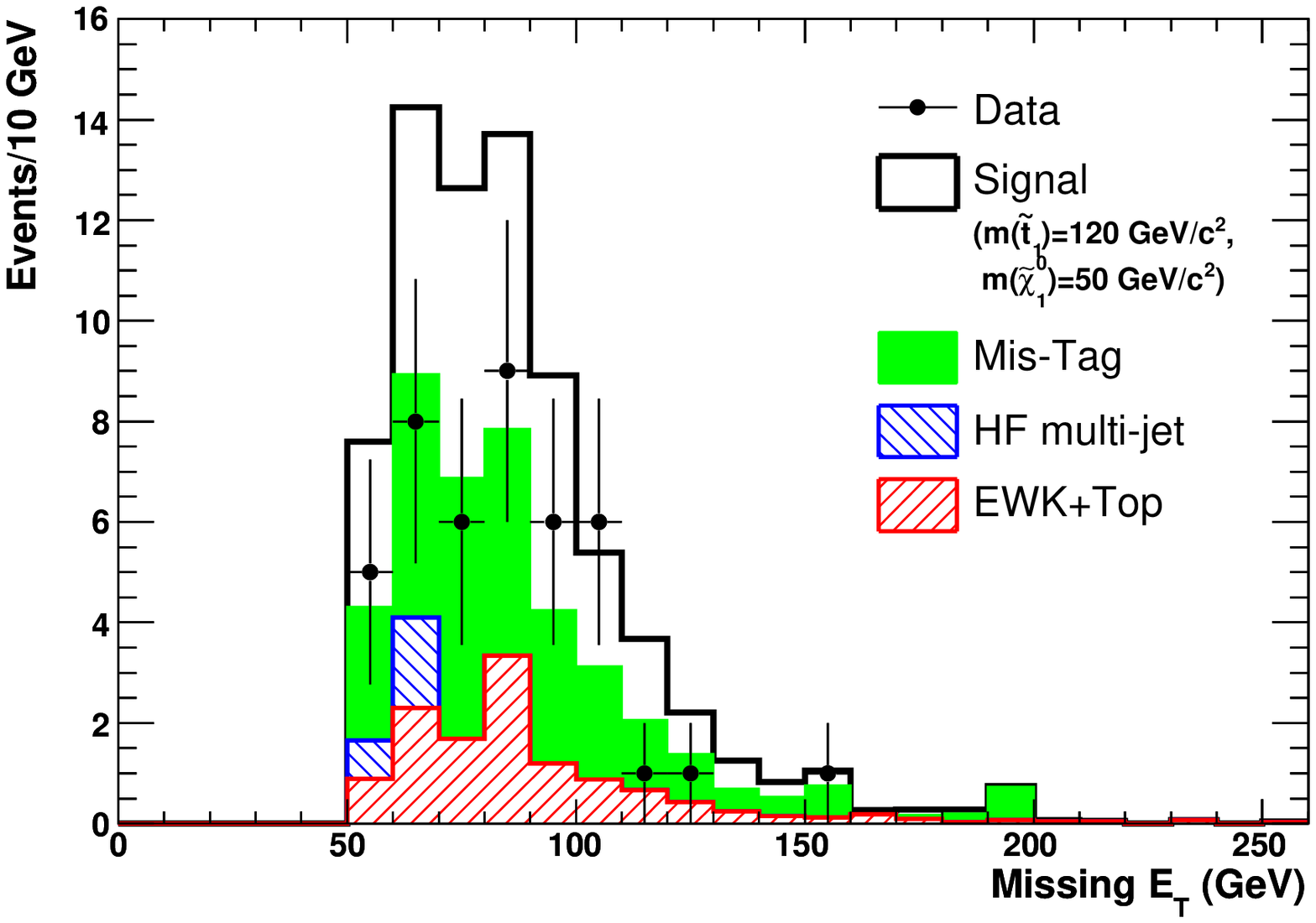}
\hfill
\includegraphics[width=9cm]{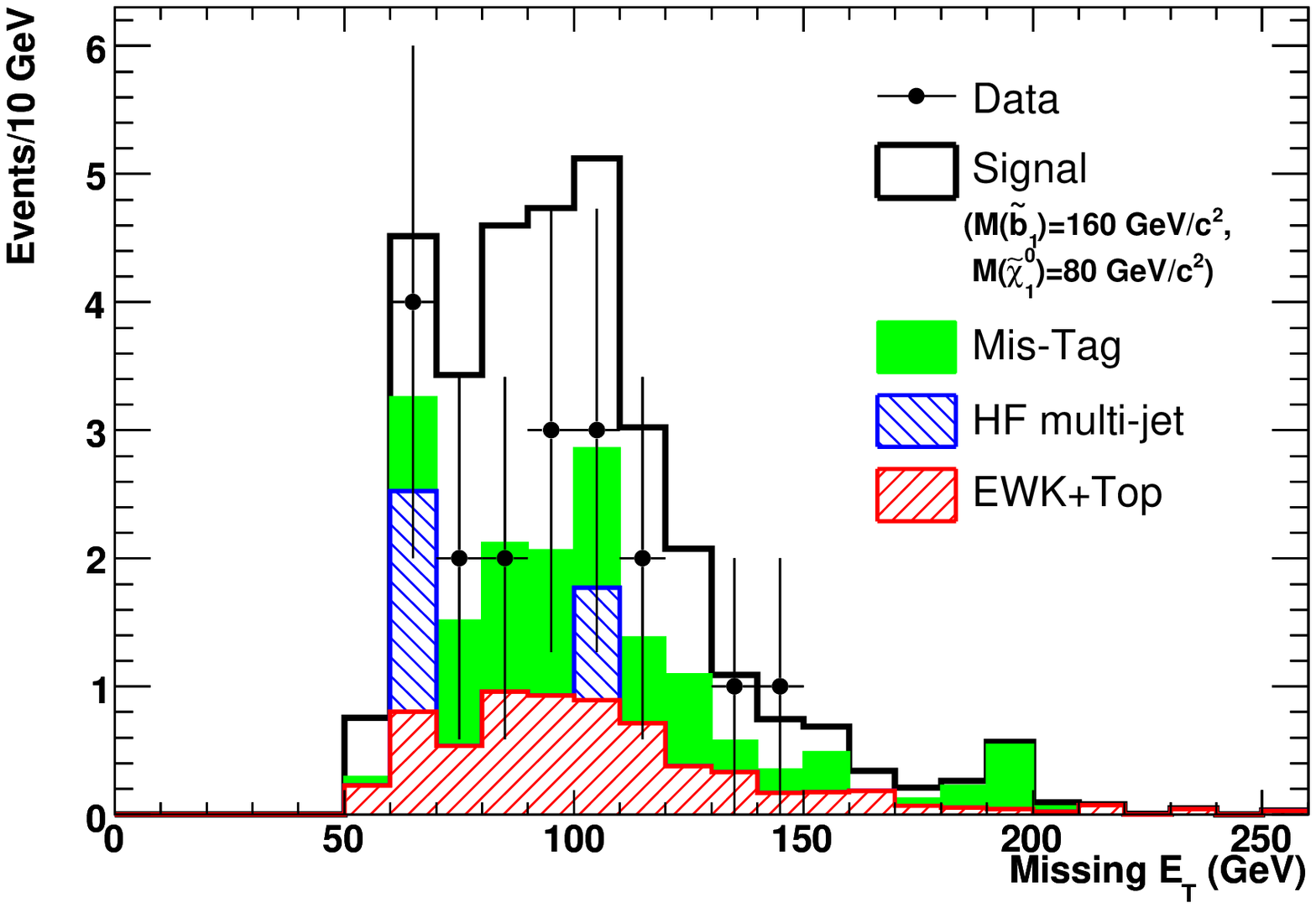}
\caption{The \metb distribution in the stop (top) and sbottom (bottom)
         signal regions for data (solid points)
         compared to the SM background (filled histograms).
         Also shown (open histograms) are the expected distributions arising from stop
         pair production and decay at $m_{\sstop_{1}}=120$ GeV/$c^{2}$ and
         $m_{\schi^{0}_{1}}=50$ GeV/$c^{2}$,
         and sbottom pair production and decay at $m_{\ssbottom_{1}}=160$ GeV/$c^{2}$
         and $m_{\schi^{0}_{1}}=80$~GeV/$c^{2}$.
        }
\label{Fig:comstopsbottom_1}
\end{figure}

We studied several kinematic distributions.
As an example Fig. \ref{Fig:comstopsbottom_1} shows
the observed \metb distributions from the data and 
the predicted \metb distributions from the SM background
for the ``high'' mass stop and ``medium'' mass sbottom searches
after all selection criteria are applied.
They are in good agreement with the distributions observed in the data.
No evidence for stop and sbottom production is observed in any of the kinematic
regions that we have studied.

%%%%%%%%%%%%%%%%%%%%%%%%%%%%%%%%%%%%
%%                                 %
%%%%%%%%%%%%%%%%%%%%%%%%%%%%%%%%%%%%

For each of the three stop/sbottom mass ranges the number of observed events
in the data is consistent with the SM expectation.
An upper limit on the possible number of signal events at
95 \% C.L. using a modified frequentist approach \cite{tjlimit}
is calculated for each mass range.
We compare this upper limit to the prediction from the NLO calculation of
{\sc prospino} using the {\sc cteq6m} PDFs.
The uncertainties on the theoretical cross section arise from the choice of
the renormalization and factorization scale and of PDFs.
The changes induced by modifying $Q_{\mbox{rf}}$ by a factor of two higher or lower than
its nominal value ($Q_{\mbox{rf}}=m_{\sstop_{1},\ssbottom_{1}}$), and the variations
observed with the {\sc cteq6m} uncertainty PDF set, result in a change of $\sim \pm 20\%$
in the theoretical cross section, when combined in quadrature.
To extract the limits on the stop and sbottom masses, we conservatively
choose the one sigma lower bound of the NLO cross section associated with
these uncertainties.

\begin{figure}[htbp]
\centering
\includegraphics[width=9cm]{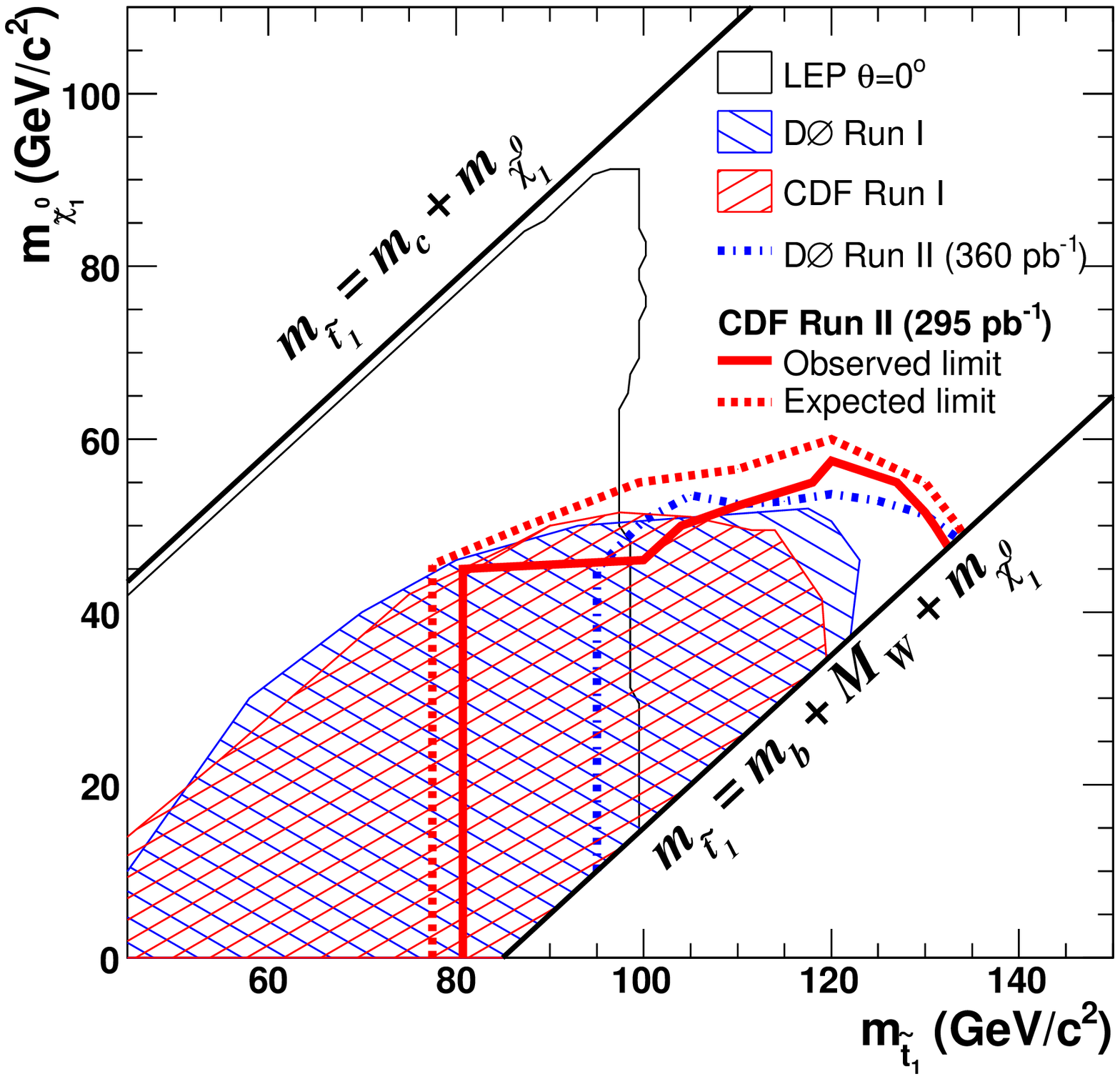}
\caption{The 95\% C.L. exclusion region in the mass plane of
$m_{\schi^{0}_{1}}$ vs $m_{\sstop_{1}}$, assuming that stop decays
exclusively into $c\schi^{0}_{1}$.
For the region excluded by the present search (area inside the solid curve),
the one sigma lower bound of the predicted NLO cross section is used to extract
the mass limits. The expected exclusion contour is shown as a dashed curve.
The result from the D\O\ Collaboration, using data from the Tevatron Run II,
is presented in the same manner \cite{D0Run2Stop}.
Results from other previous searches are also indicated
\cite{LEPStopSbottom,CDFRun1StopSbottom,D0Run1Stop}.
The LEP results are presented for the case of no mixing in the stop sector ($\theta=0^{\circ}$).}
\label{Fig:mass_plane_exclude_stop}
\end{figure}

The interpretation of the null result in the stop search is presented as a
95\% C.L. exclusion region in the mass plane of $m_{\schi^{0}_{1}}$ vs  $m_{\sstop_{1}}$,
as shown in Fig. \ref{Fig:mass_plane_exclude_stop}.
The observed exclusion is smaller than the expected exclusion because
we observe slightly more events in the data compared to the number of events
expected from the SM processes, as shown in Table \ref{tab:bgstop}.
The maximum exclusion value of $m_{\sstop_{1}}$ is 132 GeV/$c^{2}$
for $m_{\schi^{0}_{1}}=48$ GeV/$c^{2}$,
corresponding to a cross section times branching ratio limit of 2.8 pb.
The maximum $m_{\schi^{0}_{1}}$ excluded is 57 GeV/$c^{2}$ at
$m_{\sstop_{1}}=120$ GeV/$c^{2}$.
The reach in $m_{\sstop_{1}}$ is limited by the integrated luminosity,
whereas the gap between the kinematic limit
$m_{\sstop_{1}} = m_{c} + m_{\schi^{0}_{1}}$ and the
excluded region is mostly due to the \metb requirement in the event selection.

%%%%%%%%%%%%%%%%%%%%%%%%%%%%%%%%%%%%
%% FIGURE: MET (mincut)            %
%%%%%%%%%%%%%%%%%%%%%%%%%%%%%%%%%%%%

\begin{figure}[htbp]
\centering
\includegraphics[width=9cm]{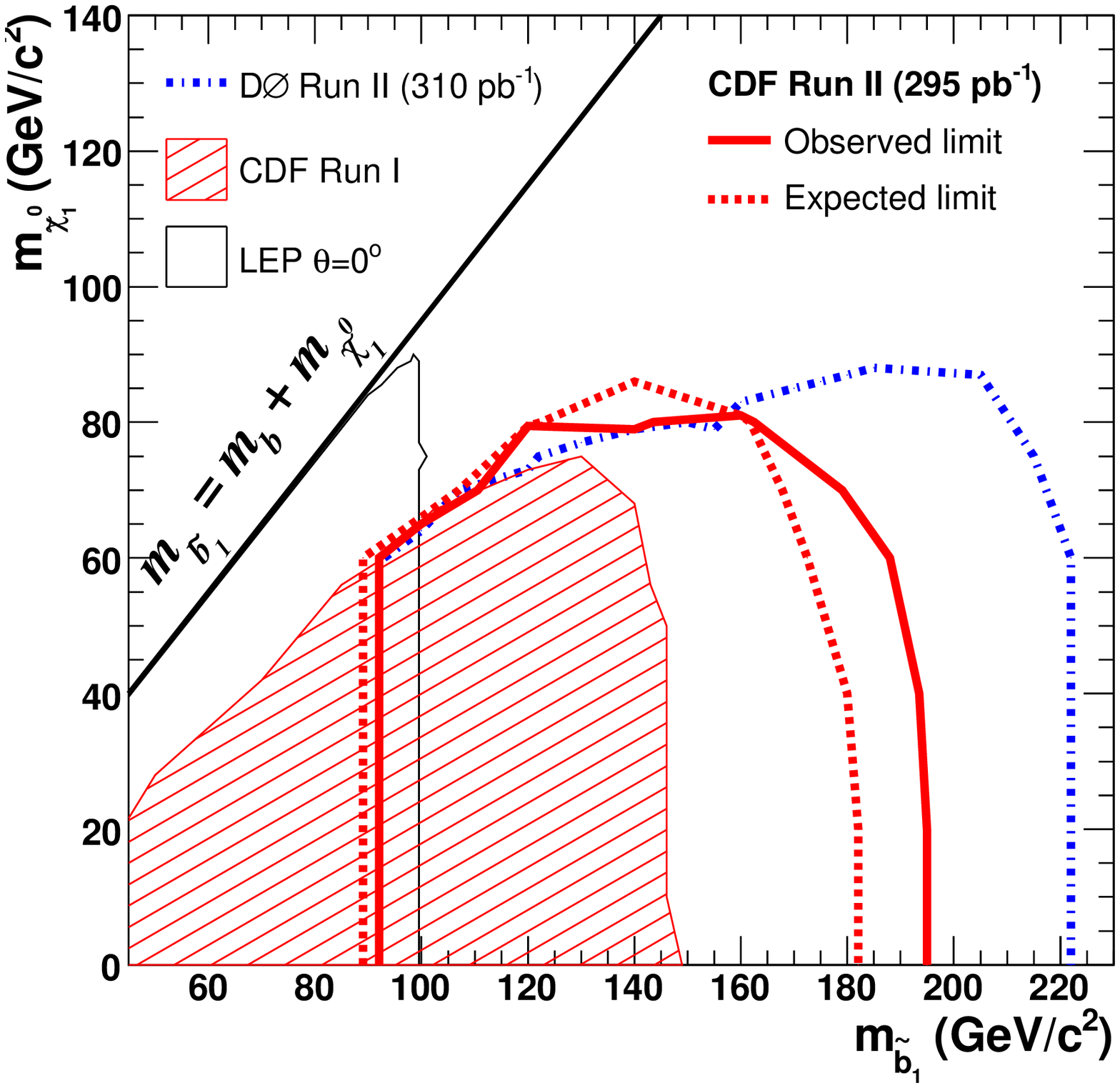}
\caption{The 95\% C.L. exclusion region in the mass plane of
$m_{\schi^{0}_{1}}$ vs $m_{\ssbottom_{1}}$,
assuming that sbottom decays exclusively into $b\schi^{0}_{1}$.
For the region excluded by the present search (area inside the solid curve),
the one sigma lower bound of
the predicted NLO cross section is used to extract the mass limits.
The dash line indicates the expected exclusion region.
The result from the D\O\ Collaboration, using data from the Tevatron Run II,
is presented in the same manner \cite{D0Run2Sbottom}.
Results from other previous searches are also indicated \cite{LEPStopSbottom,CDFRun1StopSbottom}.
The LEP results are presented for the case of no mixing in the sbottom sector ($\theta=0^{\circ}$).}
\label{Fig:mass_plane_exclude_sbottom}
\end{figure}

For the sbottom search, the interpretation of the null result
is presented as a 95\% C.L. exclusion region in the mass plane of
$m_{\schi^{0}_{1}}$ vs  $m_{\ssbottom_{1}}$,
as shown in Fig. \ref{Fig:mass_plane_exclude_sbottom}.
The observed limit is larger than the expected limit at large sbottom mass
because we observe fewer events than expected from the SM processes in the
``high'' sbottom mass range search.
The $m_{\ssbottom_{1}}$ is excluded up to 193 GeV/$c^{2}$ for
$m_{\schi^{0}_{1}}=40$ GeV/$c^{2}$.
This corresponds to a cross section times branching ratio limit of 0.25 pb.
The exclusion features of this plot are similar to those for the stop search.

%%%%%%%%%%%%%%%%%%%%%%%%%%%%%%%%%%%%
%% FIGURE: MET (mincut)            %
%%%%%%%%%%%%%%%%%%%%%%%%%%%%%%%%%%%%

In conclusion, we have performed searches for stop and sbottom production in 
proton-antiproton collisions at $\sqrt{s} = 1.96$ TeV, using 295~pb$^{-1}$ of data
recorded by the CDF experiment.
In this search we assume that the stop (sbottom) decays exclusively into
a $c$ ($b$) quark and the lightest neutralino.
The number of events that pass our selection for both stop and sbottom searches
is consistent with the standard model expectation.
Our 95\% C.L. exclusion regions in the supersymmetric quark-neutralino mass plane
extend beyond the LEP and the Tevtron's Run I reaches \cite{LEPStopSbottom,CDFRun1StopSbottom,D0Run1Stop}.
The exclusion reach in this stop search is comparable with D\O\ {\hspace{-0.3em}'s} latest
Run II stop results \cite{D0Run2Stop}.

We thank the Fermilab staff and the technical staffs of the participating institutions
for their vital contributions.
This work was supported by the U.S. Department of Energy and National Science Foundation;
the Italian Istituto Nazionale di Fisica Nucleare;
the Ministry of Education, Culture, Sports, Science and Technology of Japan;
the Natural Sciences and Engineering Research Council of Canada;
the National Science Council of the Republic of China;
the Swiss National Science Foundation;
the A.P. Sloan Foundation;
the Bundesministerium f\"ur Bildung und Forschung, Germany;
the Korean Science and Engineering Foundation and the Korean Research Foundation;
% the Particle Physics and Astronomy Research Council and the Royal Society, UK;
the Science and Technology Facilities Council and the Royal Society, UK;
the Institut National de Physique Nucleaire et Physique des Particules/CNRS;
the Russian Foundation for Basic Research; the Comisi\'on Interministerial de Ciencia y Tecnolog\'{\i}a, Spain;
the European Community's Human Potential Programme;
the Slovak R\&D Agency;
and the Academy of Finland.

%==============================================================================

%==============================================================================

\begin{thebibliography}{}
\bibitem{eta} CDF uses a cylindrical coordinate system in which $\theta$ 
is  the polar angle to the proton beam, $\phi$ is the azimuthal angle about the beam axis, and 
pseudorapidity is defined as  $\eta=-$ln tan($\theta$/2). The transverse energy and transverse momentum are defined 
as $E_{T}$ = $E~\sin\theta$ and $p_T$ = $p~\sin\theta$, where $E$ is energy measured in the calorimeter and $p$
 is momentum measured by the tracking system. The missing transverse energy vector, \vmet, 
is $-\sum_i E^i_T{\bm n}_i$, where ${\bm n}_i$ is the unit vector in the azimuthal plane that points 
from the beamline to the $i^{th}$ calorimeter tower.
The scalar quantity \met is the magnitude of the vector \vmet.
\end{thebibliography}
\end{document}